# Harmonized connectome resampling for variance in voxel sizes

Authors: Elyssa M. McMaster[a], Nancy R. Newlin[b], Gaurav Rudravaram[a], Adam M. Saunders[a], Aravind R. Krishnan[a], Lucas W. Remedios[b], Michael E. Kim[b], Hanliang Xu[b], Derek B. Archer[e], Kurt G. Schilling[c,d], François Rheault[f], Laurie E. Cutting,[a,d,g,h], and Bennett A. Landman[a,b,c,d]

[a] Department of Electrical and Computer Engineering, Vanderbilt University, Nashville, TN USA
[b] Department of Computer Science, Vanderbilt University, Nashville, TN USA
[c] Vanderbilt University Institute of Imaging Science, Vanderbilt University Medical Center, Nashville, TN USA
[d] Department of Radiology and Radiological Sciences, Vanderbilt University Medical Center, Nashville, TN USA
[e] Vanderbilt Memory and Alzheimer's Center, Vanderbilt University School of Medicine, Nashville, TN USA
[f] Department of Computer Science, Université de Sherbrooke, Sherbrooke, QC, Canada
[g] Vanderbilt Kennedy Center, Vanderbilt University, Nashville, TN USA
[h] Peabody College of Education, Vanderbilt University, Nashville, TN USA

Corresponding author: Elyssa McMaster
Email: elyssa.m.mcmaster@vanderbilt.edu

Abstract

Purpose: Diffusion MRI (dMRI) fiber tractography presents exciting opportunities to deepen our knowledge of human brain connectivity and discover novel alterations in white matter. To date, there has been no comprehensive study characterizing the effect of diffusion-weighted magnetic resonance imaging (dMRI) voxel resolution on the resulting connectome for high resolution subject data.

Methods: The statistical significance of graph measures derived from dMRI data were assessed by comparing connectomes from the same scans across different resolutions with 44 subjects (32 female) from the Human Connectome Project – Young Adult dataset (HCP-YA) with scan/rescan data (88 scans). We explored 15 isotropic and anisotropic resolutions, generated tractography and connectomes, and compared graph measures between each resolution and its nearest larger and smaller resolutions.

Results: Nearly all pairwise comparisons yielded statistically significant differences in graph measures ($p \leq 0.05$, Wilcoxon Sign-Rank Test). Upon up sampling the 14 down sampled resolutions in 0.5 mm increments, mitigation of the spatial sampling effect on both the tractography and the connectome's complex graph measures is observed. To investigate translational impact, a subject from the HCP-YA data was down sampled to the resolutions of three major national studies and up-sampled this data back to 1 mm isotropic. The Cohen's *d* effect size was at least 1 for all graph measures when comparing study resolution with 1 mm isotropic resolution data.

Conclusion: Similarity in results improved with higher resolution, even after initial down-sampling. To ensure robust tractography and connectomes, resample data to 1 mm isotropic resolution.

1. Introduction

Diffusion weighted MRI (dMRI) provides a non-invasive solution to map the random thermal motion of water molecules within biological tissue. Diffusion MRI, in combination with diffusion tensor imaging (DTI) or fiber orientation distributions (FODs), allows the mapping of the white matter bundles within the brain to further understand its function in a process called tractography [1]. In DTI, every voxel in a diffusion image is represented by a matrix and then visualized as a tensor to show the direction of water diffusion and degree of fractional anisotropy (FA) in the voxel of interest [2]. A higher voxel resolution in an image with DTI produces more tensors and can mitigate partial volume effects, but the DTI representation is limited in regions of crossing fibers [3], [4], [5]. FODs can model multiple fibers in the same voxel space, but like DTI, it is a voxel-wise operation [4], [5], [6]. The map of the streamlines of connectivity with the brain is called a tractogram, and they can be generated with either DTI or FODs [1]. The tractogram's matrix representation, a connectome, weights the brain's connections according to a metric of interest, such as number of streamlines, mean length of streamlines, or fractional anisotropy, and can be interpreted by complex graph measures [7].

Complex graph measures have been applied to many diseases and disorders, but with different image resolutions and other acquisition parameters, which can lead to variability in results. Studies of tractography and connectomes suffer from variability in their results based on sensitivity to brain parcellation [8], acquisition parameters such as angular and spatial resolution [9], [10], [11], scan-rescan variability [12], reconstruction [13], and the tractography algorithm [14]. Connectivity analysis makes it possible to look for biomarkers of disease and disorders [15]. Studies have used complex graph measures to quantify white matter changes across age groups and conditions, and they use voxel resolution ranging from state-of-the-art research scanners with high resolution acquisition resolution to clinical resolution data [16], [17], [18].

The processes for generating these tractable voxel-wise representations of the dMRI signal/human brain map are inherently tied to voxel resolution. To construct a connectome begins with the creation of the tractogram. Tractography algorithms follow anisotropy within the microns of diffusion to delineate white matter from other tissue, but the parameters to do this, like ideal voxel size and tractography step size, have no gold standard [19], [20]. Tractography involves generating seed points along the streamlines of the tractogram and generating a complex graph network. The lack of a generalized protocol for spatial resolution can lead to a wide range of results in connectomes generated from tractography, despite having the exact same anatomy, at different resolutions (Figure 1). This inconsistency introduces a harmonization problem – how do we leverage the data we have to generate more reproducible and repeatable results?

Previous studies have sought to determine the sensitivity of connectomes to other factors, but none have shown an avenue for harmonization of subject data acquired at different resolutions. Zhan, et al investigated the role of angular sampling in diffusion imaging measures [21]. Schilling, et al studied the effect of scanner protocol effects on microstructure, including acquisition resolution from different scanners [22]. Neher et. al showed the effect of anisotropic voxels in tractography with a phantom dataset [23]. In 2012, Tournier et al introduced MRTrix as a set of tools to analyze diffusion images with FODs instead of a traditional DTI model, a breakthrough in the study of tractography. Based on the 2 mm isotropic and larger spatial resolution used in most dMRI studies and a reliance on visual inspection due to a lack of metrics at the time, spatial sampling showed little effect on the tractography process [5]. More recently, connectome adjacency matrices have become a popular representation of the tractography map because they quantify metrics that may not be obvious based on visual inspection [17], [24], [25], [26]. A 2024 study by Zhong, et al collected data from two different sites with different acquisition resolutions and concluded that voxel size has a larger impact on the variation between and across subjects than scan-rescan variability or coil differences [27], which invites an exciting opportunity to leverage the tools of the last decade to understand the effect of voxel size on connectomics in a way previously imperceptible to visual inspection.

Other studies have sought to perform spatial interpolation in DTI and tractography settings. Yang, et al explored interpolation approaches in the context of DTI to improve spatial resolution of low-quality data and evaluated based on fractional anisotropy and mean diffusivity in cardiac data. They establish a relationship between different levels of interpolation and a change in fiber generation and call for a similar study in brain data [28]. Coupé et al proposed a super resolution method for diffusion weighted imaging that outperforms traditional methods like trilinear and B-spline interpolation, and showed the potential of their method by illustrating fiber tracking quality in a deterministic tractography experiment with super-resolved images that decrease voxel dimensions from 1.2 mm isotropic to 0.6 mm isotropic and 0.4 mm isotropic and relies on a b0 [29]. Super resolution in dMRI has taken off as a subject of interest in the last half-decade as the topic of the "Resolving to super resolution multi-dimensional diffusion imaging" (Super MUDI) Challenge at MICCAI 2021 [30]. Before implementing super-resolution techniques on dMRI datasets, there is a need to establish if the common problems in connectome harmonization are due to a lack of information that needs to be solved with a super-resolution approach or a fixable issue in the tractography process. If we find that we can harmonize connectomes with basic up-sampling, we can bypass the computationally expensive step of a state-of-the-art AI super resolution approach.

In this work, we aim to demonstrate spatial harmonization in the context of tractography and connectomics. We propose a study to determine the degree to which the resolution of the diffusion-weighted image impacts complex graph measures. We aim to minimize the effect of acquisition resolution for more reproducible and repeatable maps of the brain's connections. We evaluate the differences between connectomes with a pairwise comparisons between the graph measures at every resolution generated for this experiment via a Wilcoxon Signed-Rank test and compare the effect sizes between the relevant resolutions' distributions of 12 complex graph measures with the Cohen's *d* coefficient.

2. Methods

We hypothesize that the resolution of a diffusion MRI image will introduce statistically significant changes in the complex graph measures of the resulting connectomes due to the interrelated nature of spatial resolution and tractography. Additionally, we hypothesize that many of the connectome's complex network measures become more reproducible with basic up sampling (Figure 2). By regridding the image to a higher spatial resolution, we can address a common problem of tractography: multiple tissue types in a common voxel space [3]. We show that down sampling, even as little as 0.5 mm isotropic, fundamentally changes the integrity of the tractogram and the connectome because the tissue types in the voxels become more heterogenous as they increase in size. We explore the differences qualitatively with images of individual tractogaphy bundles. We visualize our methods in Figure 3.

2.1 Data

Our dataset consists of both dMRI and T1-weighted images 44 healthy subjects (32 female) of the scan/rescan data from the Human Connectome Project – Young Adult (HCP-YA) dataset. HCP used a novel diffusion imaging approach with b-values 1000, 2000, and 3000 s/mm$^2$ with a 3T customized WU-Minn-Ox HCP scanner specific to the project with a resolution of 1.25 x 1.25 x 1.25 mm$^3$, with a minimal preprocessing pipeline on to correct eddy current distortions from the diffusion gradient and to align the diffusion images to harmonize all scan modalities [31]. Average age among subjects is 30.36 years with standard deviation ± 3.34 [32].

2.2 Preprocessing

Though the diffusion data from HCP is already preprocessed, we use the PreQual preprocessing pipeline for efficient quality assurance, diffusion tensor analysis and visualization [33]. We used Freesurfer for the segmentation of the corresponding T1-weighted images for the connectome mapping and tractography bundle segmentations[34].

Once the initial HCP-YA dataset had been preprocessed with PreQual and Freesurfer, we used *mrgrid* from MRTrix3 to down sample the voxel sizes of each of the 44 scan/rescan subjects' images [35]. 14 additional datasets were generated by increasing the voxel sizes of the images in 0.5 mm increments (1.25 mm isotropic to 4.75 mm isotropic and 1.25 x 1.25 x 4.75 anisotropic). In the case of down-sampling, *mrgrid* uses Gaussian smoothing as its default interpolation method, but maintains the image's real-world coordinates. In the case of up-sampling, *mrgrid* defaults to cubic interpolation [35].

After the down sampled images were preprocessed, we simulated a study with preprocessed low-resolution data by reshaping the voxels of the lower quality images in 0.5 mm increments, isotropically and anisotropically. This allowed us to determine if the inconsistencies between low-quality and high-quality data are based on information loss or algorithmic choices because *mrgrid* only changes spatial dimensions, which introduces some degree of information loss [35].

2.3 Tractography

We generate tractograms with 10 million streamlines using MRTrix3's default probabilistic tracking of second order integration over FODs, *tckgen* [24]. This default tracking places a seed point at every half-voxel [36], again tying the quality of the tractogram to the resolution of the diffusion image. For the tractography implementation, we limited seeding and a termination using the five-tissue-type mask and allowed backtracking [24],[37]. We repeat the process with the same number of streamlines and conditions for the set of 88 HCP-YA DWI for 14 additional resolutions and the 14 resolutions of up-sampled lower resolution data.

2.4 Connectomics

The tractograms are mapped to a connectome with the Desikan-Killany atlas with 84 cortical parcellations from Freesurfer [24], [34], [38]. We generated connectomes weighted by number of streamlines, mean length of streamlines, and FA. The connectomes weighted by number of streamlines and mean length of streamlines are used to compute the complex graph measures[24], [39]. We use 12 of the graph measures commonly used to analyze connectome organization [7], [40] (see Supplemental Material for the definitions of the complex graph measures included in our study). We used Scilpy to compute the complex graph measures with Python, as defined in the Brain Connectivity Toolbox (BCT) for Matlab [41]. The node-level computations are based on all the nodes in the network averaged together [24]. Although some graph measures are dependent on each other (for example, characteristic path length and global efficiency), we share many complex graph measures for as much transparency as possible. We acknowledge that some graph analysis studies control for network properties such as density [42], but the large voxel sizes introduced in this project exhibited such instability that the control would offer little consolation (see Supplemental Material Figure 1). In a study with more voxel sizes in a closer range, this kind of analysis would offer a more robust result.

2.5 Statistical Analysis

We performed a Wilcoxon Sign-Rank test using *wilcoxon* from *scipy.stats* for our before and after measurements of the original resolution images and the down sampled images [43]. The null hypothesis asserts that the median differences between paired observations of changes in spatial resolution has no effect on the complex graph measures of a connectome, while the alternative hypothesis asserts that the differences between paired observations is not zero.

Once we rejected the null hypothesis, we up-sampled the down-sampled images to test our hypothesis about the potential to use spatial sampling for harmonization and used the Cohen's *d* coefficient to measure the effect size between the original resolution images and those with some degree of information loss from the simple down sampling applied. We used the coefficients Cohen suggests to measure effect size: $d = 0.2$ indicates small effect size; $d = 0.5$ indicates medium effect size, and $d = 0.8$ indicates large effect size [44]. We aimed to quantify the effect size of the complex graph measures of the down-sampled change in resolution to the original HCP-YA data, and then up-sample the lower quality data and compare that to the original resolution as well. We used scan/rescan effect size as a baseline for variability in the complex graph measures in our result, and then compared the first run of data between the resolutions.

### 3. Results

3.1 Quantitative Results

The result of our test showed statistical significance between nearly every resolution pair in nearly every graph measure. Preliminary experiments showed that at the step between 2.75 mm isotropic and 3.25 mm isotropic, the complex graph measures begin to exhibit large Cohen's d effect sizes between adjacent resolutions, and, qualitatively, spurious streamlines and large, unlocalized bundles. Tracts became physiologically implausible because of spatial sampling too large for a realistic human brain – this voxel resolution introduces multiple tissue types in the same space and the possibility of crossing fibers, which makes it more difficult for the probabilistic tracking algorithm to decide how to map the white matter tracts [3]. Since the voxels in major national studies do not exceed 3 mm isotropic, we decided to continue our project only regarding voxels smaller than that [32], [45], [46], [47].

The Cohen's *d* effect size between the original dataset, low resolution data, and up-sampled versions of the down-sampled data indicates an algorithmic problem to overcome instead of an information loss problem across the complex graph measures. In Figure 4, the rows with bold boarders represent the Cohen's *d* coefficients between run 1 of different resolutions, while those not marked with boarders show scan-rescan variability within a single resolution. We observe a large Cohen's *d* effect size when the data is down sampled by a factor of 0.5 mm, isotropic.

When the same 1.75mm isotropic data is up sampled to 1.25 mm isotropic, we find a smaller effect size on all measures except *clustering*.

3.2 Qualitative Results

To interpret our qualitative results and to reaffirm our quantitative results, we took one subject from the HCP-YA dataset and resampled the scan/rescan diffusion images to the resolutions with three major national studies: Alzheimer's Disease Neuroimaging Initiative (ADNI) [45], The Baltimore Longitudinal Study of Aging (BLSA) [46], and a common clinical isotropic acquisition of 2 mm isotropic [47]. By down-sampling our HCP data to the original resolutions of the major national studies, we can understand how effective the simple down-sampling strategy is for the harmonization of research data. We resampled our preprocessed 1.25 mm isotropic diffusion scan from HCP to 1.3672 x 1.3672 x 2.7 mm$^3$ for ADNI, 0.8125 x 0.8125 x 2.18856 mm$^3$ for BLSA, and 2 mm isotropic and performed all the same steps for tractography and connectomics as for the original dataset. We then took the down-sampled images from these three different resolutions and again up-sampled them to a common 1 mm isotropic, after which we performed the tractography steps again. We determined the Cohen's *d* coefficient between the three resolutions by taking the mean and standard deviation for all three original resolutions' graph measures combined and then calculating the mean and standard deviation for the three resampled datasets (see Table 1 for the complex graph measures). We found a consistent large effect size between the original data and the resampled data; the lowest absolute value of the effect size was for the participation graph measure at 1.21, much larger than 0.8. Spatial sampling played a significant role in the generation of the streamlines and calculation of the complex graph measures, and with the much smaller standard deviation measurements we observe in the high-resolution data, we see great potential to harmonize connectomes with simple changes to spatial sampling.

We use Recobundles to visualize the middle of the corpus callosum, the left arcuate fasciculus, and the right corticospinal tract [48]. We overlay the white matter bundles over the common T1-weighted image to visualize the changes in the tracts based on the resolution of the diffusion image in MI-Brain with the original dataset resolutions and the common resampled (1mm isotropic) images (Figure 5) [49]. The bundles from the major studies' resolutions show several false-positive streamlines and areas of high curvature due to the inclusion of more tissue-types in a common voxel space. The bundles that result from the common 1mm isotropic show a more concentrated body of streamlines directed toward a more specific region. The outcome of this small visualization was consistent with the rest of our study: larger voxels and lower resolution results in less repeatable, less dense streamlines.

## 4. Discussion

We have investigated the implications of the inherent relationship between diffusion MRI, tractography, connectomics, and spatial resolution, and we have articulated an exciting path forward: spatial harmonization. We found a direct connection between spatial sampling and complex graph measure variability, and we have shown that simple up-sampling on both isotropic and anisotropic voxels to a common resolution lowers the variability of tractograms and complex graph measures. By minimizing the possibility of multiple tissue types in a single voxel space while maintaining a resolution reasonable for computation, we can make the most out of the parameters of the default tracking algorithm, *tckgen*, without making any changes to the algorithmic approach. Though it is possible to change the tractogram's step size, a change in step size does not mitigate the partial volume effects that may impact the tractogram's generation and is computationally expensive. This resampling strategy shows the potential for spatial sampling harmonization and indicates an exciting first step toward its large-scale study and implementation.

Limitations

Though many complex graph measures showed decreased variability from the simple up-sampling approach, the degrees to which these measures improved are not dependent on any single measure or resolution. Specifically, the *clustering* graph measure presents a unique challenge because its effect size remains large across resolution comparisons. It is an open question as to what is driving this effect and transitivity may offer an interesting perspective. *Clustering* is a local measure that may not be as stable in reproducibility studies compared to global metrics.

Since this study only used HCP-YA data, it limits the effect of spatial sampling on individuals with white matter diseases or disorders, and of different ages. Additionally, our dataset has a bias toward female scans. This study also did not account for site differences and acquisition parameters, as all data were collected by HCP with consistent acquisition parameters. We believe that these variations may add additional complex graph measure variability that simple up-sampling may not be able to account for. We also acknowledge the possible biases in our connectome computation that could result from our choice of the relatively coarse Desikan-Killiany atlas [38]. Variations in graph measures may occur with an implementation of a finer parcellation [50]. We note that resampling higher than 1 mm isotropic may be extremely computationally intensive for any downstream data processing.

We found statistical significance at every 0.5 mm step, isotropic and anisotropic, in the complex graph measures of the HCP-YA data and its lower-resolution counterparts. With simple up-sampling, we produce more reproducible and repeatable complex graph measures, even with the information loss inherent to down sampling. We have shown that this strategy is effective on resolutions used in major national studies on white matter diseases and disorders. Our recommendation is to up sample diffusion data to 1 mm isotropic before performing DTI and tractography to minimize effects of spatial sampling on the process.

Conflicts of Interest

The authors have no conflicts of interest to declare.

CRediT Statement

Elyssa McMaster: *Conceptualization, Methodology, Software, Validation, Formal Analysis, Investigation, Data Curation, Writing – Original Draft, Writing – Review & Editing, Visualization*
Nancy Newlin: *Conceptualization, Methodology, Software, Validation, Formal Analysis, Investigation, Writing – Review & Editing, Visualization, Supervision*
Gaurav Rudravaram: *Software, Validation, Writing – Review & Editing*
Adam M. Saunders: *Formal Analysis, Writing – Review & Editing, Visualization*
Aravind R. Krishnan: *Formal Analysis, Writing – Review & Editing, Visualization*
Lucas W. Remedios: *Formal Analysis, Writing – Review & Editing, Visualization*
Michael E. Kim: *Software, Data Curation, Writing – Review & Editing*

Hanliang Xu: *Conceptualization, Data Curation, Writing – Review & Editing*

Derek B. Archer: *Conceptualization, Methodology, Writing – Review & Editing*

Kurt G. Schilling: *Conceptualization, Data Curation, Writing – Review & Editing*

François Rheault: *Conceptualization, Software, Validation, Writing – Review & Editing*

Laurie E. Cutting: *Writing – Review & Editing, Supervision, Funding Acquisition*

Bennett A. Landman: *Conceptualization, Methodology, Validation, Formal Analysis, Resources, Writing – Review & Editing, Visualization, Supervision, Project Administration, Funding Acquisition*

Acknowledgements

We would like to acknowledge Dr. Mikail Rubinov for his feedback in the preparation of this work. The Vanderbilt Institute for Clinical and Translational Research (VICTR) is funded by the National Center for Advancing Translational Sciences (NCATS) Clinical Translational Science Award (CTSA) Program, Award Number 5UL1TR002243-03. This work was conducted in part using the resources of the Advanced Computing Center for Research and Education at Vanderbilt University, Nashville, TN. ADSP U24AG074855, NIH 1R01EB017230 Uncertainty in Diffusion MRI. This work was supported by Integrated Training in Engineering and Diabetes, grant number T32 DK101003 and National Cancer Institute (NCI), Grant/Award Number: R01 CA253923. The content is solely the responsibility of the authors and does not necessarily represent the official views of the NIH.

During the preparation of this work the author(s) used ChatGPT in order to create code segments based on task descriptions, as well as debug, edit, and autocomplete code. After using this tool/service, the author(s) reviewed and edited the content as needed and take(s) full responsibility for the content of the publication.

Figures

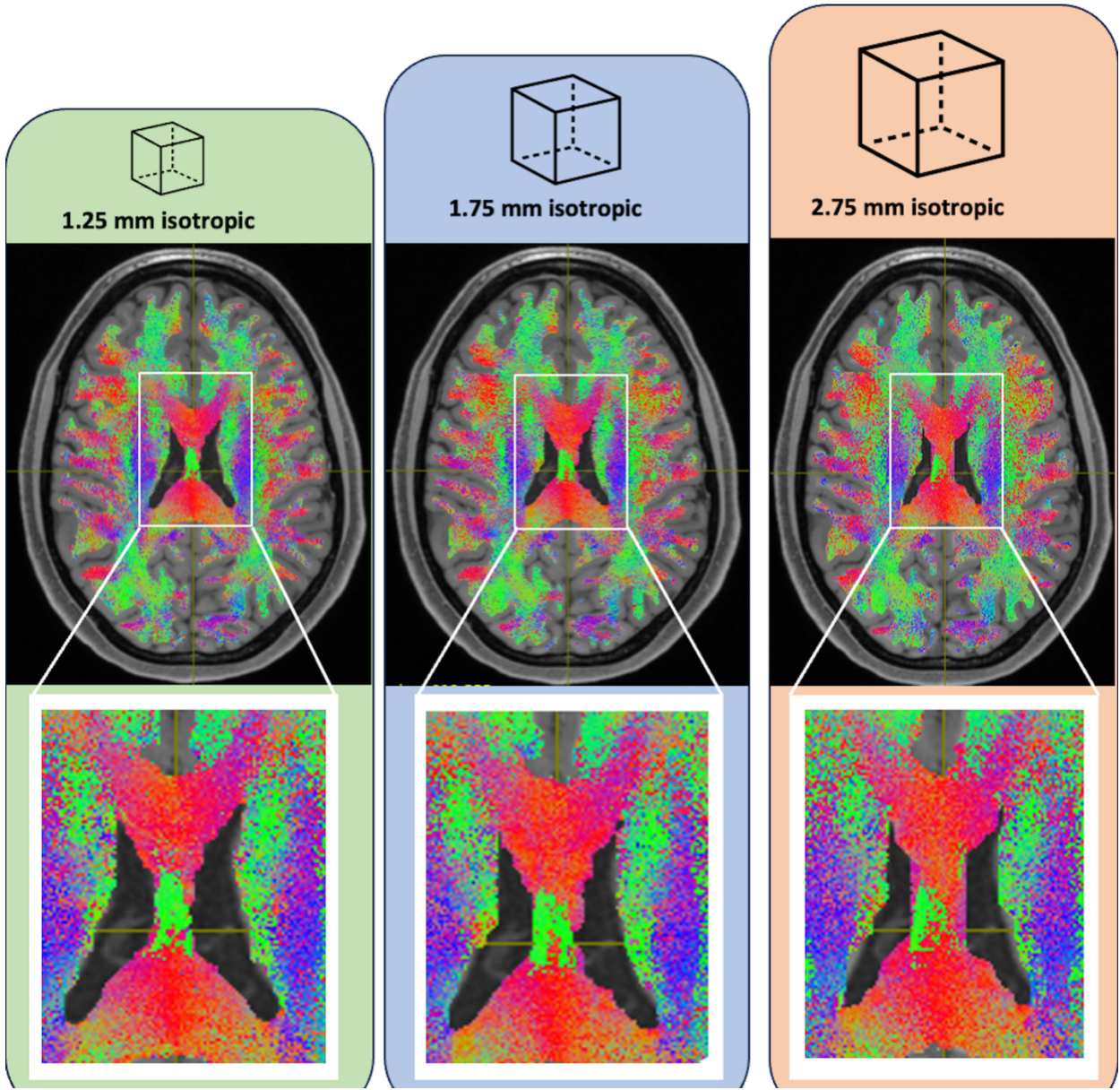

Figure 1: We analyze one subject at the acquisition resolution of 1.25 mm isotropic, and down sampled to 1.75 mm isotropic and 2.75 mm isotropic. Though the images show the same anatomy in the same acquisition, please note the thickening of the corpus callosum and the narrowing of the ventricles as the voxels increase in size.

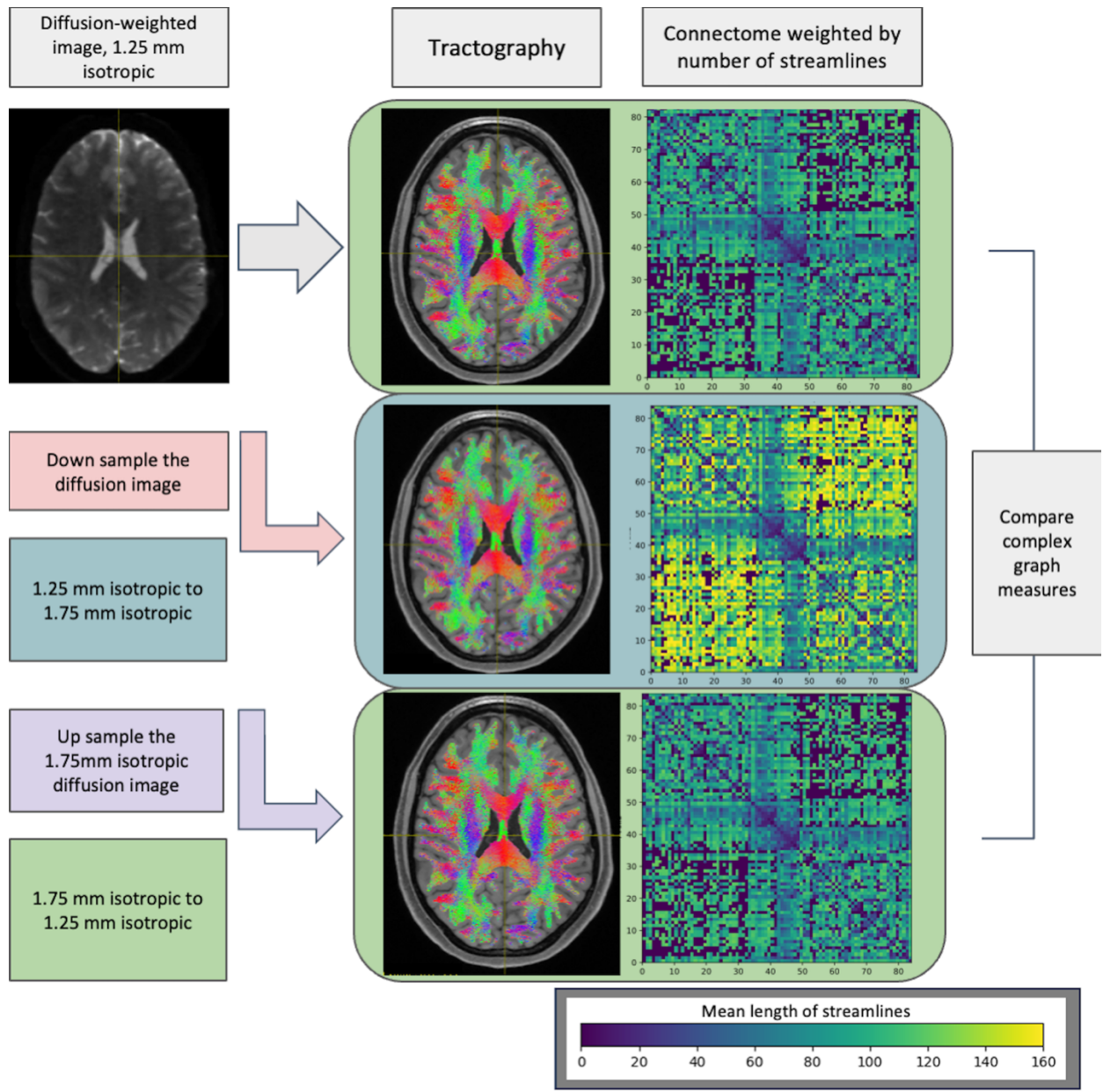

Figure 2: We aim to characterize the effect of spatial sampling on tractography and connectomes with a study that uses the same scans to down sample and subsequentially up sample the same scans. This study design allows us to understand both the effect of down sampling and low voxel resolution on the tractogram and analyze which graph measures are restored or deteriorated when a lower resolution image is up sampled. The down-sampled image may result in longer streamlines between regions due to false positive or spurious streamlines.

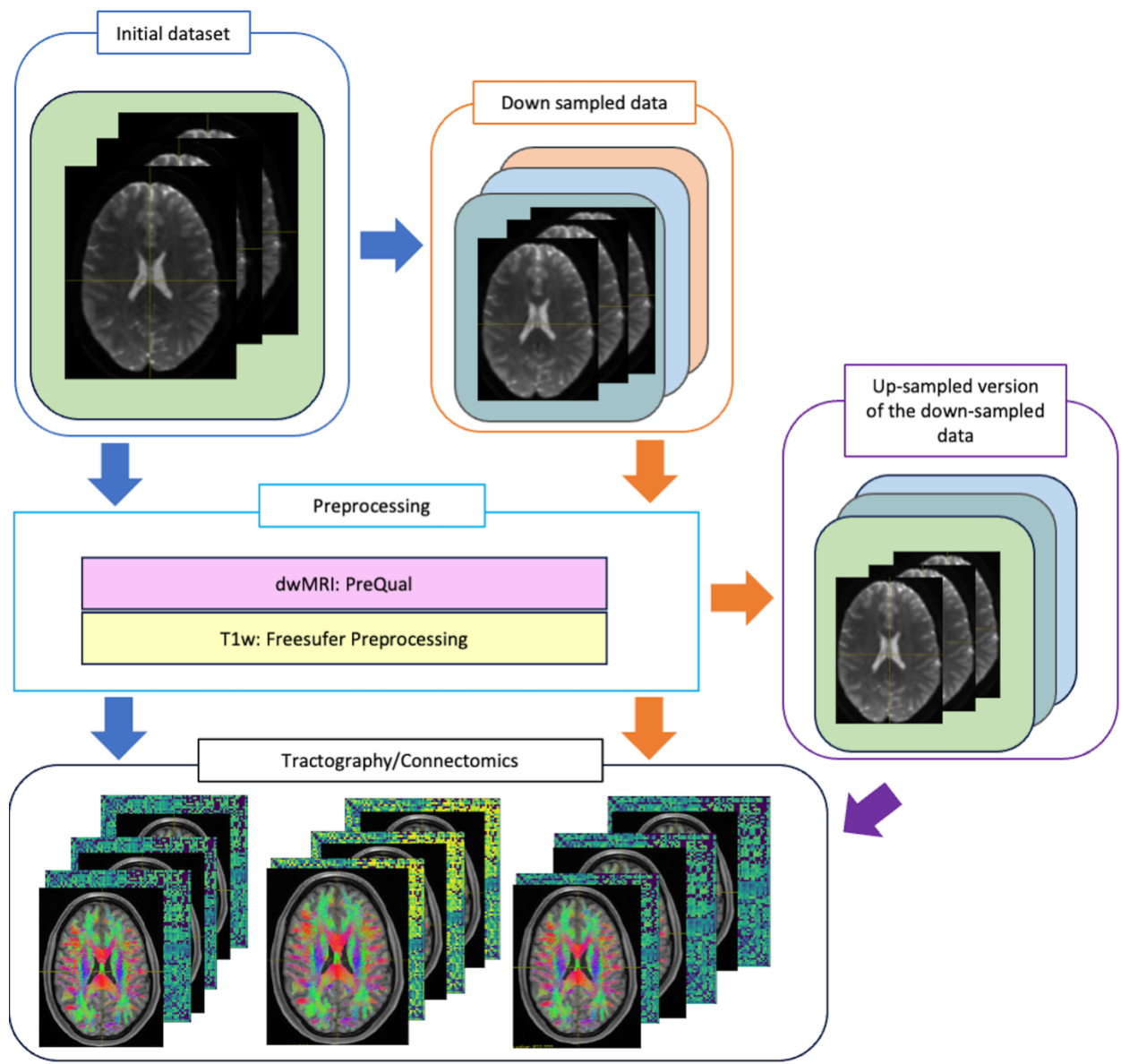

Figure 3: We designed the method to both understand the changes that happen between low- and high-resolution data and to take a low-quality dataset and preprocess and up sample it the way one would if they only had low quality data. This way, we can understand the impact of information loss on lower quality scans.

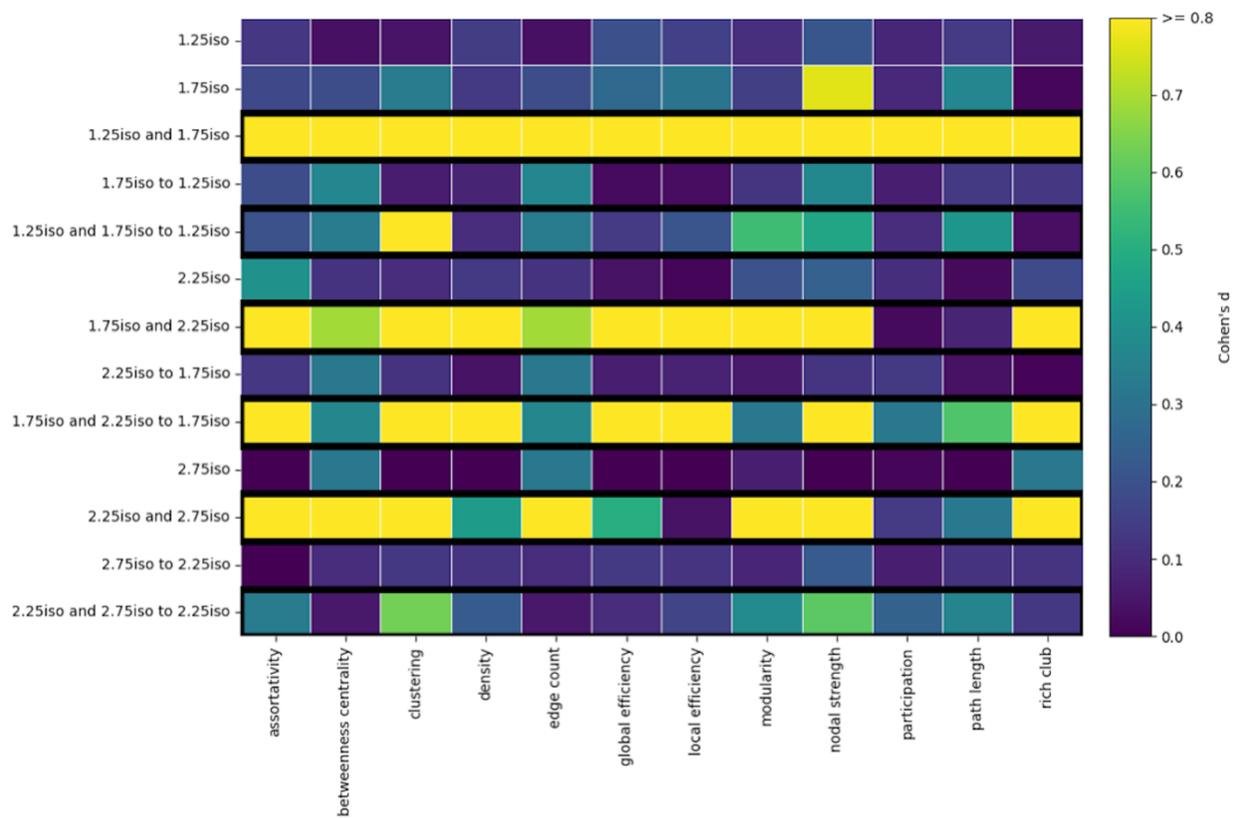

Figure 4: We used scan/rescan variability as a baseline to quantify the effect size inherent to the tractography process and then compared run 1 of different resolutions to measure the effect size of spatial resolution. We found that reshaping the voxels on lower quality data made the graph measures in their connectomes to be more like the original high-quality data than larger-voxel counterparts, which leads to a smaller effect size and more repeatable data. The comparisons between run 1 of two different resolutions are indicated with the bold outlined boxes, and the scan-rescan comparisons within a single resolution are shown as default.

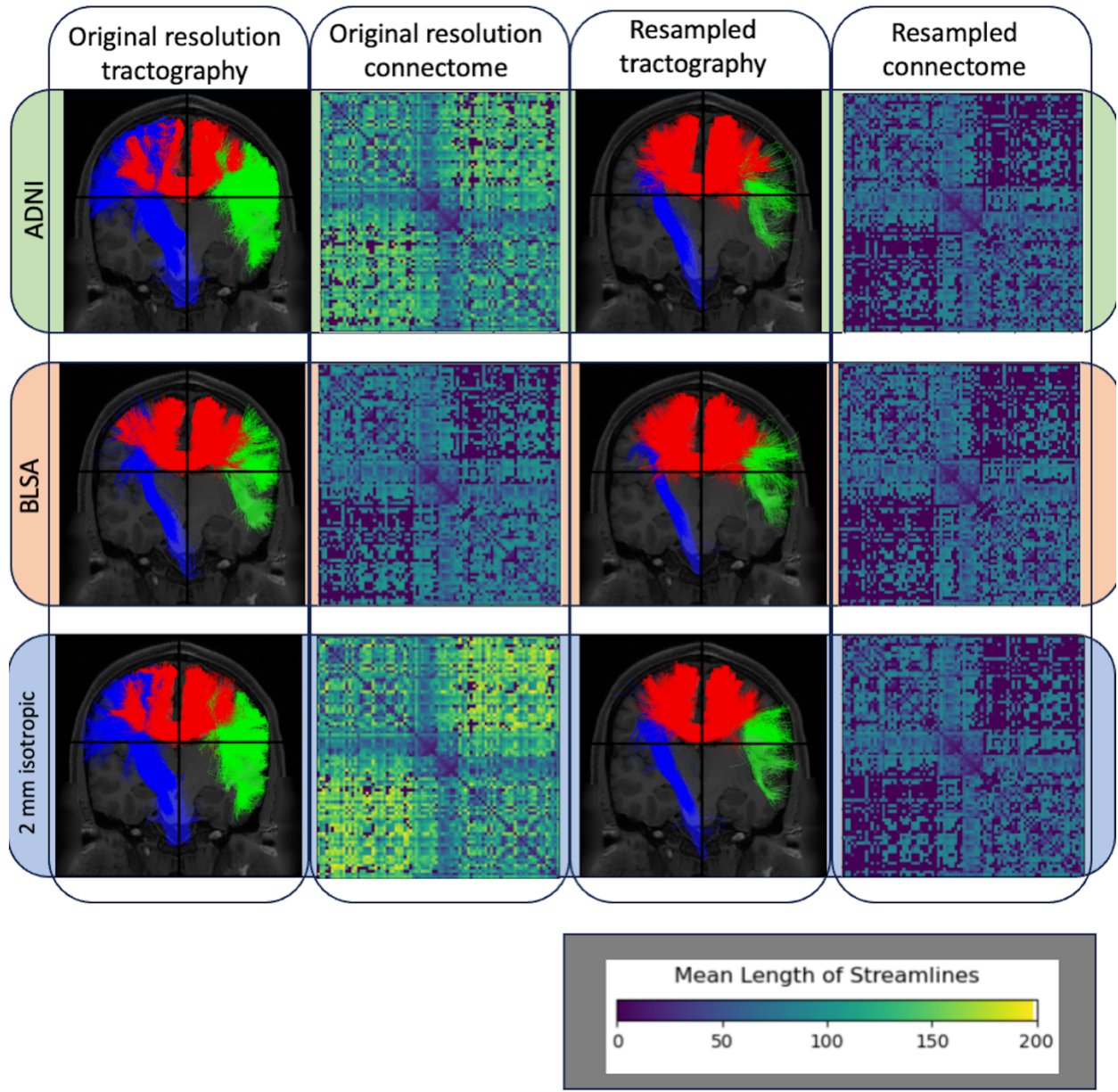

Figure 5: In the anterior view of the middle corpus callosum, left arcuate fasiculus, and right corticospinal tract, we see the effect of voxel size variation on the same subject with high curvature, loss of localization, and enlargement of the tracts. We observe wide variation in the length of streamlines between the regions of the Desikan-Killiany atlas. When the same diffusion

image is down sampled from the study's resolution, the tractography appears more localized and anatomically plausible.

|  | ADNI, original resolution | ADNI, resampled to 1mm isotropic | BLSA, original resolution | BLSA, resampled to 1 mm isotropic | 2 mm isotropic, original resolution | 2 mm isotropic, resampled to 1mm isotropic |
|---|---|---|---|---|---|---|
| assortativity | -0.019 | -0.002 | 0.001 | 0.007 | -0.028 | 0.009 |
| betweenness centrality | 0.025 | 0.022 | 0.024 | 0.023 | 0.023 | 0.024 |
| clustering | 530.231 | 574.099 | 522.250 | 560.984 | 576.773 | 573.194 |
| density | 0.938 | 0.674 | 0.748 | 0.671 | 0.988 | 0.664 |
| edge count | 3.007 | 2.786 | 2.896 | 2.836 | 2.873 | 2.909 |
| global efficiency | 107.043 | 63.952 | 73.703 | 63.739 | 117.647 | 63.263 |
| local efficiency | 97.884 | 64.187 | 72.046 | 64.034 | 105.949 | 63.204 |
| modularity | 0.603 | 0.697 | 0.680 | 0.703 | 0.575 | 0.704 |
| nodal strength | 172193.726 | 157612.155 | 160342.357 | 156332.893 | 177235.190 | 156186.345 |
| participation | 0.319 | 0.289 | 0.288 | 0.294 | 0.377 | 0.285 |
| path length | 72.649 | 61.476 | 63.403 | 61.791 | 77.660 | 57.979 |
| rich club | 0.807 | 0.437 | 0.517 | 0.428 | 0.882 | 0.426 |

Table 1: The complex graph measures for the HCP-YA subject resampled to the resolutions of the major national studies and then up-sampled to 1mm isotropic.

Citations

[1] S. Farquharson *et al.*, "White matter fiber tractography: Why we need to move beyond DTI," *J Neurosurg*, vol. 118, no. 6, 2013, doi: 10.3171/2013.2.JNS121294.

[2] C. F. Westin, S. E. Maier, H. Mamata, A. Nabavi, F. A. Jolesz, and R. Kikinis, "Processing and visualization for diffusion tensor MRI," *Med Image Anal*, vol. 6, no. 2, 2002, doi: 10.1016/S1361-8415(02)00053-1.

[3] K. G. Schilling *et al.*, "Prevalence of white matter pathways coming into a single white matter voxel orientation: The bottleneck issue in tractography," *Hum Brain Mapp*, vol. 43, no. 4, pp. 1196–1213, Mar. 2022, doi: 10.1002/hbm.25697.

[4] J. D. Tournier, "Diffusion MRI in the brain – Theory and concepts," *Progress in Nuclear Magnetic Resonance Spectroscopy*, vol. 112–113. 2019. doi: 10.1016/j.pnmrs.2019.03.001.

[5] J. D. Tournier, F. Calamante, and A. Connelly, "MRtrix: Diffusion tractography in crossing fiber regions," *Int J Imaging Syst Technol*, vol. 22, no. 1, 2012, doi: 10.1002/ima.22005.

[6] "dwi2response." Accessed: May 06, 2024. [Online]. Available: https://mrtrix.readthedocs.io/en/dev/reference/commands/dwi2response.html

[7] M. Rubinov and O. Sporns, "Complex network measures of brain connectivity: Uses and interpretations," *Neuroimage*, vol. 52, no. 3, pp. 1059–1069, Sep. 2010, doi: 10.1016/j.neuroimage.2009.10.003.

[8] S. Qi, S. Meesters, K. Nicolay, B. M. ter Haar Romeny, and P. Ossenblok, "The influence of construction methodology on structural brain network measures: A review," *Journal of Neuroscience Methods*, vol. 253. Elsevier B.V., pp. 170–182, Sep. 01, 2015. doi: 10.1016/j.jneumeth.2015.06.016.

[9] S. Crater *et al.*, "Resolution and b value dependent structural connectome in ex vivo mouse brain," *Neuroimage*, vol. 255, Jul. 2022, doi: 10.1016/j.neuroimage.2022.119199.

[10] A. Guidon, A. Batrachenko, A. V. Avram, and A. W. Song, "The Effect of Spatial Resolution on Structural Connectome Analysis."

[11] K. S. Ambrosen *et al.*, "Validation of structural brain connectivity networks: The impact of scanning parameters," *Neuroimage*, vol. 204, Jan. 2020, doi: 10.1016/j.neuroimage.2019.116207.

[12] M. J. Vaessen, P. A. M. Hofman, H. N. Tijssen, A. P. Aldenkamp, J. F. A. Jansen, and W. H. Backes, "The effect and reproducibility of different clinical DTI gradient sets on small world brain connectivity measures," *Neuroimage*, vol. 51, no. 3, pp. 1106–1116, Jul. 2010, doi: 10.1016/j.neuroimage.2010.03.011.

[13] K. Mori, I. Sakuma, Y. Sato, C. Barillot, and N. Navab, Eds., *Medical Image Computing and Computer-Assisted Intervention – MICCAI 2013*, vol. 8149. in Lecture Notes in Computer Science, vol. 8149. Berlin, Heidelberg: Springer Berlin Heidelberg, 2013. doi: 10.1007/978-3-642-40811-3.

[14] M. Bastiani, N. J. Shah, R. Goebel, and A. Roebroeck, "Human cortical connectome reconstruction from diffusion weighted MRI: The effect of tractography algorithm," *Neuroimage*, vol. 62, no. 3, pp. 1732–1749, Sep. 2012, doi: 10.1016/j.neuroimage.2012.06.002.

[15] A. Zalesky, A. Fornito, and E. T. Bullmore, "Network-based statistic: Identifying differences in brain networks," *Neuroimage*, vol. 53, no. 4, pp. 1197–1207, Dec. 2010, doi: 10.1016/j.neuroimage.2010.06.041.

[16] W. Yuan, S. L. Wade, and L. Babcock, "Structural connectivity abnormality in children with acute mild traumatic brain injury using graph theoretical analysis," *Hum Brain Mapp*, vol. 36, no. 2, pp. 779–792, Feb. 2015, doi: 10.1002/hbm.22664.

[17] M. Daianu *et al.*, "Rich club analysis in the Alzheimer's disease connectome reveals a relatively undisturbed structural core network," *Hum Brain Mapp*, vol. 36, no. 8, pp. 3087–3103, Aug. 2015, doi: 10.1002/hbm.22830.

[18] E. L. Dennis *et al.*, "Changes in anatomical brain connectivity between ages 12 and 30: A HARDI study of 467 adolescents and adults," in *2012 9th IEEE International Symposium on Biomedical Imaging (ISBI)*, IEEE, May 2012, pp. 904–907. doi: 10.1109/ISBI.2012.6235695.

[19] B. Jeurissen, M. Descoteaux, S. Mori, and A. Leemans, "Diffusion MRI fiber tractography of the brain," *NMR Biomed*, vol. 32, no. 4, Apr. 2019, doi: 10.1002/nbm.3785.

[20] M. S. Pinto *et al.*, "Harmonization of Brain Diffusion MRI: Concepts and Methods," *Frontiers in Neuroscience*, vol. 14. Frontiers Media S.A., May 06, 2020. doi: 10.3389/fnins.2020.00396.

[21] L. Zhan *et al.*, "How does angular resolution affect diffusion imaging measures?," *Neuroimage*, vol. 49, no. 2, pp. 1357–1371, Jan. 2010, doi: 10.1016/j.neuroimage.2009.09.057.

[22] K. G. Schilling *et al.*, "Fiber tractography bundle segmentation depends on scanner effects, vendor effects, acquisition resolution, diffusion sampling scheme, diffusion sensitization, and bundle segmentation workflow," *Neuroimage*, vol. 242, p. 118451, Nov. 2021, doi: 10.1016/j.neuroimage.2021.118451.

[23] P. Neher, B. Stieltjes, I. Wolf, H. Meinzer, and K. H. Maier-Hein, "Analysis of tractography biases introduced by anisotropic voxels," 2013. [Online]. Available: https://www.researchgate.net/publication/236022704

[24] N. R. Newlin, F. Rheault, K. G. Schilling, and B. A. Landman, "Characterizing Streamline Count Invariant Graph Measures of Structural Connectomes," *Journal of Magnetic Resonance Imaging*, vol. 58, no. 4, 2023, doi: 10.1002/jmri.28631.

[25] M. P. van den Heuvel and O. Sporns, "Rich-club organization of the human connectome," *Journal of Neuroscience*, vol. 31, no. 44, pp. 15775–15786, Nov. 2011, doi: 10.1523/JNEUROSCI.3539-11.2011.

[26] P. Imms *et al.*, "The structural connectome in traumatic brain injury: A meta-analysis of graph metrics," *Neurosci Biobehav Rev*, vol. 99, pp. 128–137, Apr. 2019, doi: 10.1016/j.neubiorev.2019.01.002.

[27] J. Zhong *et al.*, "Quantitative Brain Diffusion Metrics are Fragile to Voxel Size: A Prospective Volunteer Study Calling for Protocol Standardization." [Online]. Available:

https://submissions.mirasmart.com/ISMRM2024/Itinerary/Files/PDFFiles/ViewAbstract.aspx

[28] F. Yang, Y. M. Zhu, M. Robini, and P. Croisille, "DT-MRI interpolation: At what level?," in *International Conference on Signal Processing Proceedings, ICSP*, 2012. doi: 10.1109/ICoSP.2012.6491592.

[29] P. Coupé, J. V. Manjón, M. Chamberland, M. Descoteaux, and B. Hiba, "Collaborative patch-based super-resolution for diffusion-weighted images," *Neuroimage*, vol. 83, 2013, doi: 10.1016/j.neuroimage.2013.06.030.

[30] V. Nath *et al.*, "Resolving to super resolution multi-dimensional diffusion imaging (Super-MUDI)," May 2021.

[31] M. F. Glasser *et al.*, "The minimal preprocessing pipelines for the Human Connectome Project," *Neuroimage*, vol. 80, 2013, doi: 10.1016/j.neuroimage.2013.04.127.

[32] M. F. Glasser *et al.*, "The Human Connectome Project's neuroimaging approach," *Nature Neuroscience*, vol. 19, no. 9. Nature Publishing Group, pp. 1175–1187, Sep. 01, 2016. doi: 10.1038/nn.4361.

[33] L. Y. Cai *et al.*, "PreQual: An automated pipeline for integrated preprocessing and quality assurance of diffusion weighted MRI images," *Magn Reson Med*, vol. 86, no. 1, pp. 456–470, Jul. 2021, doi: 10.1002/mrm.28678.

[34] B. Fischl, "FreeSurfer," *NeuroImage*, vol. 62, no. 2. pp. 774–781, Aug. 15, 2012. doi: 10.1016/j.neuroimage.2012.01.021.

[35] "mrgrid." Accessed: May 06, 2024. [Online]. Available: https://mrtrix.readthedocs.io/en/dev/reference/commands/mrgrid.html

[36] "tckgen", Accessed: May 08, 2024. [Online]. Available: https://mrtrix.readthedocs.io/en/dev/reference/commands/tckgen.html

[37] N. Newlin *et al.*, "Learning site-invariant features of connectomes to harmonize complex network measures," SPIE-Intl Soc Optical Eng, Feb. 2024, p. 92. doi: 10.1117/12.3009645.

[38] "CorticalParcellation." Accessed: May 05, 2024. [Online]. Available: https://surfer.nmr.mgh.harvard.edu/fswiki/CorticalParcellation

[39] "tck2connectome." Accessed: May 06, 2024. [Online]. Available: https://mrtrix.readthedocs.io/en/dev/reference/commands/tck2connectome.html

[40] O. Sporns *et al.*, "Brain Connectivity Toolbox."

[41] SCIL, "scil_connectivity_graph_measures.py." Accessed: May 08, 2024. [Online]. Available: https://github.com/scilus/scilpy/blob/f95bd6df3262fc50ef512c93dbde7e6e80276658/scripts/scil_connectivity_graph_measures.py#L7

[42] S. L. Simpson and P. J. Laurienti, "Disentangling Brain Graphs: A Note on the Conflation of Network and Connectivity Analyses," *Brain Connectivity*, vol. 6, no. 2. 2016. doi: 10.1089/brain.2015.0361.

[43] "scipy.stats.wilcoxon." Accessed: May 06, 2024. [Online]. Available: https://docs.scipy.org/doc/scipy/reference/generated/scipy.stats.wilcoxon.html

[44] J. Cohen, "Statistical Power Analysis for the Behavioral Sciences Second Edition."

[45] C. R. Jack *et al.*, "The Alzheimer's Disease Neuroimaging Initiative (ADNI): MRI methods," *Journal of Magnetic Resonance Imaging*, vol. 27, no. 4. 2008. doi: 10.1002/jmri.21049.

[46] R. J. O'Brien *et al.*, "Neuropathologic studies of the Baltimore longitudinal study of aging (BLSA)," *Journal of Alzheimer's Disease*, vol. 18, no. 3. 2009. doi: 10.3233/JAD-2009-1179.

[47] D. B. Archer *et al.*, "Leveraging longitudinal diffusion MRI data to quantify differences in white matter microstructural decline in normal and abnormal aging," *Alzheimer's and Dementia: Diagnosis, Assessment and Disease Monitoring*, vol. 15, no. 4, 2023, doi: 10.1002/dad2.12468.

[48] E. Garyfallidis *et al.*, "Recognition of white matter bundles using local and global streamline-based registration and clustering," *NeuroImage*, vol. 170. 2018. doi: 10.1016/j.neuroimage.2017.07.015.

[49]  F. Rheault, J.-C. Houde, N. Goyette, F. C. Morency, and M. Descoteaux, "MI-Brain, a software to handle tractograms and perform interactive virtual dissection." [Online]. Available: https://itk.org/

[50]  A. Zalesky *et al.*, "Whole-brain anatomical networks: Does the choice of nodes matter?," *Neuroimage*, vol. 50, no. 3, 2010, doi: 10.1016/j.neuroimage.2009.12.027.

[51]  M. E. J. Newman, "Assortative Mixing in Networks," *Phys Rev Lett*, vol. 89, no. 20, 2002, doi: 10.1103/PhysRevLett.89.208701.

[52]  L. C. Freeman, "A Set of Measures of Centrality Based on Betweenness," *Sociometry*, vol. 40, no. 1, 1977, doi: 10.2307/3033543.

[53]  J. Saramäki, M. Kivelä, J. P. Onnela, K. Kaski, and J. Kertész, "Generalizations of the clustering coefficient to weighted complex networks," *Phys Rev E Stat Nonlin Soft Matter Phys*, vol. 75, no. 2, 2007, doi: 10.1103/PhysRevE.75.027105.

[54]  SCIL, "connectivity_tools.py." Accessed: May 19, 2024. [Online]. Available: https://github.com/scilus/scilpy/blob/master/scilpy/connectivity/connectivity_tools.py

[55]  M. E. J. Newman, "Modularity and community structure in networks," *Proc Natl Acad Sci U S A*, vol. 103, no. 23, 2006, doi: 10.1073/pnas.0601602103.

[56]  R. Guimerà and L. A. N. Amaral, "Functional cartography of complex metabolic networks," *Nature*, vol. 433, no. 7028, 2005, doi: 10.1038/nature03288.

[57]  D. J. Watts and S. H. Strogatz, "Collective dynamics of 'small-world' networks," in *The Structure and Dynamics of Networks*, vol. 9781400841356, 2011. doi: 10.1515/9781400841356.301.